\documentclass[12pt,onecolumn]{emulateapj}
\usepackage{epsf}
\usepackage{epsfig}
\usepackage{amssymb,amsmath}
\def\plotonef#1{\centering\leavevmode\epsfxsize=0.5\textwidth\epsfbox{#1}}
\def\plotone#1{\centering\leavevmode\epsfxsize=\textwidth\epsfbox{#1}}

\newcommand{\beq}{\begin{equation}}
\newcommand{\eeq}{\end{equation}}
\newcommand{\bea}{\begin{eqnarray*}}
\newcommand{\eea}{\end{eqnarray*}}

\begin{document}

\title{The effect of Primordial Black Holes and streaming motions
on structure formation.}

\author{F. Atrio-Barandela\altaffilmark{1}}
\altaffiltext{1}{F\'{\i}sica Te\'orica. Universidad de Salamanca.
37008 Salamanca, Spain; atrio@usal.es}

\begin{abstract}

Primordial Black Holes  could be an important component of the dark matter 
in the Universe. If they exist, they would add a Poisson component 
to the matter power spectrum. The extra power would speed up the emergence
of dark matter halos that seed the formation of first stars and galaxies. 
Kashlinsky (2021) suggested that the
additional velocity fluctuations would accelerate the infall of baryons onto the
dark matter potential wells. We analyze the effect of
Primordial Black Holes on the baryon infall from recombination to reionization 
and find the correction to be a few percent of the power suppression
first identified by Tseliakhovich \& Hirata (2010). However, the
dynamical effect of this correction in addition to
the extra power speeds up the formation
of halos in the mass range of $10^4-10^{5-6}$M$_\odot$, while slightly
decreasing the formation of those in the range $10^6-10^8$M$_\odot$
confirming earlier analytic estimates and recent results of numerical 
simulations.
\end{abstract}


\section{Introduction.}

The detection of gravitational waves (GWs) by the LIGO 
and VIRGO  observatories (Abbott et al. 2016, 
Abbott et al. 2020) originating from the merger of binary black
hole (BH) systems revived the idea that Primordial Black Holes (PBHs)
could constitute a large fraction, if not all, of the dark matter (DM) in the 
Universe (Bird et al. 2016, Clesse \& Garc{\'\i}a-Bellido 2016,
Kashlinsky 2016, Carr, Clesse \& Garc{\'\i}a-Bellido 2021).
Such PBHs could have originated during the QCD epoch (Jedamzik 1997).
The detected binary BH mergers were unlikely to be composed
of BHs with large spins aligned to the orbital angular momentum as
one could expect from stellar BHs (Abbott et al. 2019).
Particularly interesting were those PBHs in the mass range 
$M_{PBH}\sim 10-50$M$_\odot$ as they could be the source of the binary BH events
detected by LIGO/VIRGO. The event rate can be fitted by a combined 
population of PBHs and stellar BHs mergers (Franciolini et al. 2022)
but models where all the DM is composed of PBHs
also match the LIGO/Virgo observed merger rates (Jedamzik 2021).
In parallel, astrophysical constraints have limited the mass range and fraction
of PBHs (Lacki \& Beacom 2010,
Ali-Ha\"{\i}moud \& Kamionkowski 2017, Poulin et al. 2017, 
Oguri et al. 2018, Laha 2019, Blaineau, Moniez \& Afonso 2022) 
and current estimates in the mass range
of interest limit the fraction of PBHs to be $f_{PBH}\sim 10^{-4}-0.1$
(Carr \& K\"uhnel 2020) although those constrains could be weaken
if PBHs are clustered (Belotsky et al 2019) and/or 
have an extended mass function. Carr et al. (2021) argued that a 
multi-modal PBH mass spectrum would make compatible $f_{PBH}\sim 1$
with astronomical constrains.  
Kashlinsky, Ali-Ha\"{\i}moud, Clesse et al. (2019) discussed
the prospects that new instruments and observations in the coming years could probe the 
existence of PBHs and determine their fraction in the DM component.

A population PBHs would add a Poisson isocurvature 
component to the adiabatic matter density perturbations 
generated during inflation. This contribution would dominate the power at small scales
(Meszaros 1975, Afshordi, McDonald \& Spergel 2003), increasing 
the abundance of early dark matter halos. At each redshift, the increment 
on the number density of collapsing halos would result in a
more efficient formation of first stars and galaxies, helping to
explain the level of source-subtracted cosmic infrared background (CIB) fluctuations 
(Kashlinsky 2016) that cannot 
be accounted for by the known galaxy populations (Kashlinsky et al. 2005, 2007).
Furthermore, baryon accretion onto the population of PBHs would 
contribute to various cosmic backgrounds and explain, for instance,
the coherence of the X-ray and near-infrared backgrounds 
(Cappelluti et al. 2013, Cappelluti et al. 2017, see 
Kashlinsky et al. 2018 for a review).  Detailed studies
of the emission over cosmic time from star formation and black hole accretion in a 
PBH-DM universe have shown the viability of this type of models
(Hasinger, 2020; Cappelluti, Hasinger \& Natarajan 2021). 

In this work we will consider the effect of a PBHs on the evolution of 
matter density perturbations and the formation of DM halos during reionization.
After matter radiation equality, DM density perturbations grow
while baryons and photons are still tightly coupled; later, at recombination,
baryons fall into the DM potential wells. The baryonic sound
speeds falls to $\sim 6$ km s$^{-1}$ while the DM moves at $\sim 30$ km s$^{-1}$.
Tseliakhovich \& Hirata (2010, hereafter TH) noticed 
that the supersonic motion of baryons relative to the DM rest frame 
suppressed the abundance of halos. Baryons do not fall
into the DM potential wells as fast as they would without the 
velocity effect resulting in linear fluctuations in the matter density 
being suppressed.
The scales most affected were $k\sim 200h/$Mpc and the effect was soon
confirmed by numerical simulations (O'Leary \& McQuinn, 2012). 
The effect of streaming velocities (advection) on the formation of DM halos
and stellar objects has been extensively considered (Maio, Koopmans \& Ciardi 2011,
Tseliakhovich, Barkana \& Hirata 2011, Richardson, Scannapieco
\& Thacker 2013)
concluding that the formation of halos in the mass range $(10^4,10^8)$M$_\odot$
were suppressed. Stacy, Bromm \& Loeb (2011) also identified that
when a halo is formed it does
not contain a gas core as dense as it would have without advection.
Nevertheless, the subsequent collapse of the gas 
and the formation of Pop III stars was very similar to the no-streaming case.
Kashlinsky (2021) was the first to point out 
that PBHs would lead to a more efficient redistribution 
of kinetic energy between baryons and DM on the scales relevant for the formation 
of the first sources but did not quantify the amplitude of this effect. The details 
of the baryonic infall process are of great importance for understanding 
star formation and the evolution of the interstellar medium at redshifts $z>10$. 
The reionization epoch could be directly probed through the absorption of the 
21cm line (see Barkana 2016 for a review) and indirectly through the residual 
CIB fluctuations in source subtracted maps in the near infrared (NIR) 
(Kashlinsky et al. 2018). Both probes could, consequently, 
test the effect of streaming velocities and PBHs. If the latter
play an important role on the collapse of early halos
then the physical effects during reionization generated by this population 
could be observable with forthcoming data 
(Kashlinsky, Arendt, Cappelluti et al. 2019, Hasinger 2020, Cappelluti et al. 2021).

Numerical simulations have only recently started to 
consider the effect of streaming motions and PBHs on the formation
of first stars. Liu, Zhang \& Bromm (2022) compared the results of
two simulations with streaming motions, with and without PBHs
and found that the effect of streaming motions were weaker with PBHs.
In this article we analyze how PBHs alter the evolution of matter 
density perturbations in the presence of streaming motions. 
We will consider $f_{PBH}\sim 10^{-3}-1$ and an effective mass of
$M_{PBH}=30$M$_\odot$.  In Sec.~2 we summarize
the TH formalism. In Sec.~3 we present our main results and discuss their
implications and in Sec.~4 we summarize our main conclusions.

\section{Formalism.}

On scales of a few Mpc, the DM (assumed to be cold dark matter, CDM) 
moves with a bulk velocity of $\sim 30$km/s when the baryon sound speed 
is $\sim 6$km/s.  The velocity difference is generated by large scale modes 
of wavelength up to $\sim 200$Mpc. On scales smaller than $\sim 1$Mpc 
baryons are subject to a uniform and supersonic  motion (Mach number 
${\cal M} \sim 5$) in the DM rest frame.  These regions are the background 
where smaller scale density perturbations grow due to gravitational instability. 
TH showed that in this scenario the equations describing the evolution of 
matter perturbations at first order contain non-linear terms that change the
conclusions of linear theory.

Newtonian theory accurately describes
the evolution of subhorizon density perturbations of
baryons and DM in the matter dominated regime, when 
fluctuations in the radiation field can be ignored (Peebles 1980).
In the presence of uniform streaming motions, 
the background motion of baryons and DM can be separated
from the peculiar velocity of small scale perturbations
by writing $\vec{v}_j(\vec{x},t)=\vec{v}_j^{(bg)}(t) +\vec{u}_j(\vec{x},t)$,
where the super index $(bg)$ denotes the background (streaming) motion 
between baryons and DM within the patch.
If PBHs were present in the early Universe, they will affect the
velocity field by adding an isocurvature component to the DM power spectrum. 
In order to apply the
TH formalism we need to demonstrate that baryons and PBHs are still moving 
coherently with respect to each other on patches of a few Mpc. 

Let $\delta_j=(\rho_j-\bar{\rho}_j)/\bar{\rho}_j$ denote the density 
contrast of a matter component and $\vec{v}_j$ its peculiar velocity;
$j=(b,c)$ where $(b)$ refers to baryons and $(c)$ to DM. 
Peculiar velocities are related to matter density perturbations as
\beq
\vec{v}_j(\vec{k})=
-i\frac{H}{(1+z)}\frac{d\ln D}{d\ln z}\delta_j(\vec{k},t)\frac{\vec{k}}{k^2} ,
\eeq 
where density perturbations are written as $\delta_j(\vec{k},z)=
D_j(z)\delta_j(\vec{k})$. 
The growth factor $D_j(z)$ gives the time evolution of density perturbation of
baryons and DM; $H=H_0E(z)$ and  $E^2(z)=\Omega_{m}(1+z)^3+\Omega_{r}(1+z)^4+\Omega_\Lambda$.
$H_0,\Omega_{m},\Omega_{r},\Omega_\Lambda$ are the current values of the Hubble
constant, the matter and radiation energy densities and the energy
density associated with the cosmological constant $\Lambda$.
Let us denote the logarithmic growth by $F_j=d\ln D_j/d\ln z$. The
variance of the streaming velocity field between
baryons and dark matter can be written as
\beq
v_{bc}^2=\langle \vec{v}_{bc}\cdot\vec{v}_{bc}\rangle=
\int_0^\infty\frac{k^3|\vec{v}_{bc}|^2}{2\pi^2}\frac{dk}{k}=
\int_0^\infty\Delta^2_{v,bc}(k)d\ln k .
\label{eq:v_streaming}
\eeq
The square fluctuation streaming velocity power per logarithmic 
k-interval is
\beq
\Delta^2_{v,bc}(k)=\frac{1}{2\pi^2}\frac{H^2}{(1+z)^2}k(F_b\delta_b-F_c\delta_c)^2.
\label{eq:Delta_v}
\eeq
It gives the contribution to the variance of the streaming velocity field
per unit interval in $\ln k$. $F_b,F_c$ are the logarithmic growth functions for 
baryons and DM, respectively. 

PBHs add an extra contribution to the matter
power spectrum due to the fluctuation on the
number density of PBHs. The power spectrum is constant, independent of 
scale and is given by (Afshordi et al. 2003)
\begin{equation}
P_{PBH}(k)=\frac{9}{4}\left(\frac{1+z_{eq}}{D_+(z)}\right)^2\frac{M_{PBH}}{f_{PBH}\rho_{cr}
\Omega_{c}}\approx 1.26\times 10^{-8}
\left(\frac{1330}{D_+(1020)}\right)^2
\left(\frac{M_{PBH}}{30M_\odot}\right)\left(\frac{1}{f_{PBH}}\right)
\left(\frac{0.125}{\Omega_ch^2}\right)
{\rm Mpc}^{-3}
\end{equation}
where $z_{eq}\approx 3413$ is the redshift of matter radiation equality,
$D_+(z)$ is the linear growth factor of matter density perturbations
from redshift $z$ till today normalized to unity at present,
$M_{PBH}$ is the effective mass of PBHs, $f_{PBH}$ is the fraction of DM that is composed
of PBHs, $\rho_{cr}$ is the critical density and $\Omega_c$
is the fraction of the energy density that is made of DM particles
(including PBHs). 

\begin{figure}
\plotonef{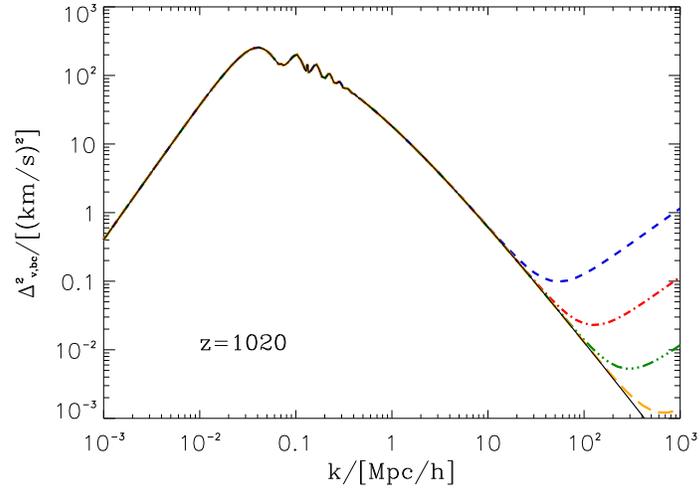}\bigskip
\caption{\small Variance of the streaming velocity field between baryons and DM. 
The dashed (blue), dot-dashed (red), triple dot-dashed (green) and long dashes (gold)
lines also include the contribution from PBHs with fractions
$f_{PBH}=(1,0.1,0.01,10^{-3})$, respectively.  The solid black line
corresponds to the concordance $\Lambda$CDM model, i.e., $f_{PBH}=0$.
}
\label{fig1}
\end{figure}

Fig.~\ref{fig1} plots the contribution of the different scales 
to the variance of the velocity difference
between baryons and DM. The solid (black) line represents
the streaming velocity variance, given by eq.~(\ref{eq:Delta_v}), of the concordance
$\Lambda$CDM model with densities $\Omega_c=0.255$, $\Omega_b=0.0455$,
Hubble parameter $h=0.7$ and spectral index at large scales $n_s=0.98$.
The dashed (blue), the dot-dashed (red), the
triple dot-dashed (green) and the long dashed (gold) lines corresponds
to PBHs fractions $f_{PBH}=1,0.1,0.01,10^{-3}$, respectively.
Plots correspond to redshift $z=1020$. In the matter dominated regime
the logarithmic growth factors are
$F_c\approx F_b\approx 1$. However, the contribution of the radiation energy density
to the background expansion is not negligible near decoupling.
For a more accurate treatment, we obtained the logarithmic growth 
factors using
the baryon and CDM transfer functions at redshifts $z_1=1020$ and $z_2=1010$,
computed using the CMBFAST code (Seljak \& Zaldarriaga, 1996).
Fig.~\ref{fig1} demonstrates that the streaming velocity is still dominated by 
contributions coming from scales in the range $k\in[0.01,1]h$Mpc$^{-1}$.
The effect of PBHs depends on $f_{PBH}$ and it is only relevant
at $k\ge 50h$Mpc$^{-1}$. Beyond this scale, the constant matter power spectrum
from PBHs results on the streaming velocity variance growing as 
$\Delta^2_{v,bc}(k)\propto k$. It could be expected that for scales $k\ge 1h$kpc$^{-1}$
the random motion induced by PBHs would disrupt
the uniform supersonic motion of baryons in the DM rest frame.
Numerical simulations have shown that the
presence of PBHs gives rise to a significant nonlinear effect that it 
is not present in the standard $\Lambda$CDM model, damping 
the matter power spectrum compared with the expectation from linear theory
(see Fig~4 of Inman \& Ali-Ha\"{\i}moud 2019).  By $z=100$, the 
matter power spectrum is damped on scales $k\ge 0.2h$kpc$^{-1}$ when 
the fraction of PBHs is larger than $f_{PBH}\ge 0.1$.
Therefore, the homogeneity of the streaming baryon flow in the
DM rest frame will not be disrupted. For $f_{PBH}<0.1$, the evolution 
is well described by linear theory but the contribution of PBHs at 
$k\ge 1h$kpc$^{-1}$ is much smaller than the contribution from $k\ge 1h$Mpc$^{-1}$. 
In summary, the random motions induced by the PBHs power spectrum
will not cancel the effect of DM-baryon supersonic motion
and the decomposition of the velocity field into a uniform background motion
and a peculiar velocity field at small scales can be applied for any fraction of PBHs.

The equations of evolution baryons and DM density perturbations of a mode 
with wavenumber $\vec{k}$ and in the matter dominated regime 
are given by (TH, eqs. 11)
\begin{equation}\label{eq:adv}
\begin{split}
\frac{\partial\delta_c}{\partial t}&=\frac{i}{a}\vec{v}^{(bg)}_{bc}\cdot\vec{k}\delta_c-\theta_c\\
\frac{\partial\theta_c}{\partial t}&=\frac{i}{a}\vec{v}^{(bg)}_{bc}\cdot\vec{k}\theta_c-
\frac{3}{2}H^2(\Omega_c\delta_c+\Omega_b\delta_b)-2H\theta_c\\
\frac{\partial\delta_b}{\partial t}&= -\theta_b\\
\frac{\partial\theta_b}{\partial t}&=-\frac{3}{2}H^2(\Omega_c\delta_c+\Omega_b\delta_b)-2H\theta_c
+\frac{c_S^2k^2}{a^2}\delta_b
\end{split}
\end{equation}
where $\theta$ is the divergence of the peculiar velocity field $\vec{u}_j$, given by
$\vec{u}_j=-ia(\vec{k}/k^2)\theta(\vec{k})$. These equations are
written in the baryon rest frame but they could have been equally written in the DM 
rest frame (O'Leary \& McQuinn 2012). TH solved these equations assuming
perturbations evolved in a background where baryon and DM densities were 
equal to the mean. On these regions the background velocities would
be constant in position but their amplitudes would decrease in time with the scale 
factor $a(t)$ as $\vec{v}_b^{(bg)},\vec{v}_c^{(bg)} \sim a^{-1}$. Their relative 
velocity would scale similarly, $\vec{v}_{bc}^{(bg)}= (\vec{v}_b^{(bg)}-
\vec{v}_c^{(bg)})\sim a^{-1}$.  Ahn (2016) showed that the relative velocity 
$\vec{v}^{(bg)}_{bc}$ scaled as the inverse of the scale factor even
if the baryon and DM densities of the background were not the mean density,
so we will use eqs.~(\ref{eq:adv}) from TH instead of a more general treatment.  
The effect of the bulk motion in eqs.~(\ref{eq:adv}) is encoded in the terms containing
the streaming velocity factor $\vec{v}^{(bg)}_{bc}$, since if $\vec{v}^{(bg)}_{bc}=0$
eqs.~(\ref{eq:adv}) would simply describe the evolution of matter density perturbation 
in the linear regime. 

\section{Numerical Results and discussion.} 

\begin{figure}
\centering 
\plotone{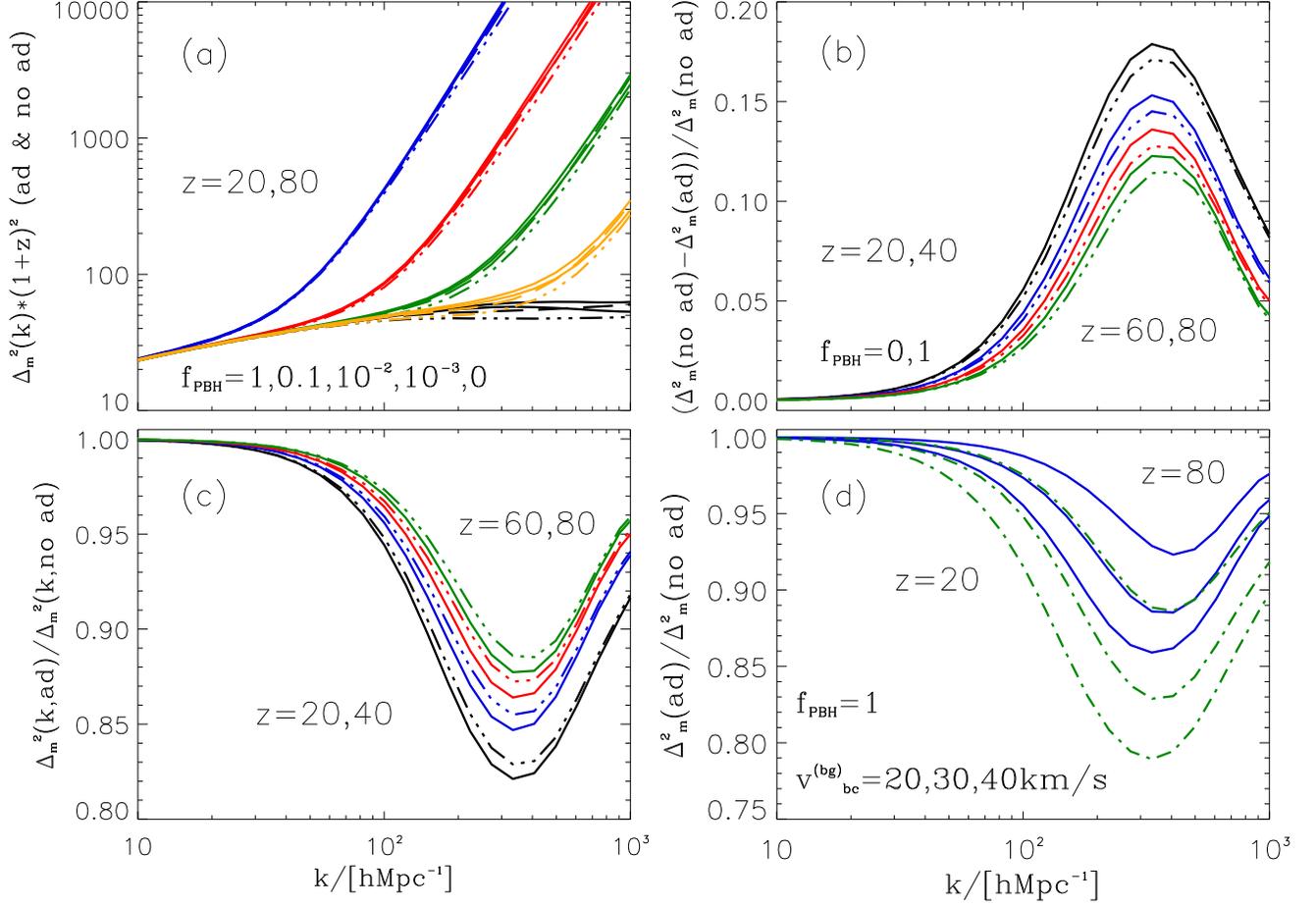}\bigskip
\caption{\small (a) Variance of the matter distribution times its
redshift dependence. From left to right, blue, green, red, gold and black lines 
correspond to $f_{PBH}=1,0.1,10^{-2},10^{-3},0$. Solid lines correspond to 
no-advection and broken lines to advection. Dashed lines and their nearby solid lines
correspond to $z=80$ while the triple dot-dashed lines and nearby solid lines
correspond to $z=20$. In (b) we plot the relative difference of the variance.
Solid and triple dot-dashed lines correspond to $f_{PBH}=0$ and $f_{PBH}=1$, respectively.
From top to bottom lines correspond to $z=20$ (black), $z=40$ (blue), $z=60$
(red) and $z=80$ (green). In (c) we plot the ratio of the variance with
advection to variance without advection. Lines follow the same convention than
in (b). In (d) we show the effect of varying the background advection
velocity. The top three solid blue lines correspond to $z=80$ and
the bottom green dot-dashed ones to $z=20$. At each redshift, the top, middle
and bottom lines correspond to $v_{bc}^{(bg)}=20,30,40$km/s, respectively.
}
\label{fig2}
\end{figure}

We solved numerically eqs.~(\ref{eq:adv}) to determine the contribution of 
PBHs to the advection of small scale perturbations by the DM streaming motions.
At redshift $z=1020$ the r.m.s streaming velocity is $v_{bc}^{(bg)}\approx 30km/s$;
adding the PBHs component increases this amplitude by a 0.1\%. 
For a better understanding of the effect of PBHs, we fixed  $v_{bc}^{(bg)}=30$km/s 
in both the standard model and in models including PBHs. The numerical solution 
will depend on the  cosine of the angle between the direction of the bulk flow velocity 
$\vec{v}^{(bg)}_{bc}$ and the wavenumber $\vec{k}$. 
We follow TH and define the isotropic average fluctuation power per
logarithmic interval, or variance, as
\beq
\Delta^2_m(k)=\frac{1}{2}\int_{-1}^{1}\Delta^2_m(k,\cos\theta)d\cos\theta
\label{eq:6}
\eeq
where $\Delta_m^2(k)=k^3P(k)/2\pi^2$ and $P(k)=|\delta_m(k)|^2$ is the matter power 
spectrum that includes 
the baryon, DM and PBHs components, i.e, $\delta_m=\delta_c+\delta_b+\delta_{PBH}$.
$\Delta^2_m(k,\cos\theta)$ is the same magnitude for each angle $\theta$ between the background 
bulk flow velocity $\vec{v}_{bc}^{(bg)}$ and  $\vec{k}$.
We computed the effect at redshifts $z=20,40,60$ and $80$ and the results 
are presented in Fig.~\ref{fig2}.
In the matter regime $\Delta_m$ scales linearly with redshift
in the absence of advection. In Fig.~\ref{fig2}a we plot $\Delta^2_m(k)*(1+z)^2$ 
to show the effect of advection at different redshifts. To avoid clutter, we plot only
$z=20,80$ that bracket the variation with redshift. Lines correspond to 
$f_{PBH}=1,0.1,10^{-2},10^{-3},0$ from left to right and are shown with
blue, green, red, gold and black lines, respectively. Solid lines correspond
to no-advection and broken lines to advection. The triple dot-dashed line
corresponds to $z=20$ and the dashed line to $z=80$. The figure shows
that at any redshift PBHs increase the power
at small scales and their effect on advection is second order; 
in most cases the differences can not be seen. 
In Fig.~\ref{fig2}b we plot the relative differences between the variances.
Solid lines correspond to $f_{PBH}=0$ and the triple dot-dashed lines
to $f_{PBH}=1$. Again, to avoid clutter we do not plot the result for
other fractions since they will fall between the two cases represented in the plot.
From top to bottom, lines correspond to $z=20$ (black), $z=40$ (blue), 
$z=60$ (red) and $z=80$ (green). 
When all other parameters are kept fixed, 
the relative effect of PBHs is a few percent that, combined with the
added power, gives rise to significant differences in the number density of
objects as we shall discuss in the next section.
In  Fig.~\ref{fig2}c we represent the variance ratio
of advection to no-advection. From top to bottom, pairs of lines correspond to
$z=80,60,40$ and $20$, coded by color as green, red, blue and black, respectively. Solid lines
correspond to no-advection and triple dot-dashed lines to advection with $f_{PBH}=1$.
All other fractions fall between the two plotted cases. 
Unity indicates that advection does not affect the evolution
of density perturbations.  

The advection velocity is, on average, $v_{bc}^{(bg)}=30$km/s, but it
will vary from one patch to the next. In 
Fig.~\ref{fig2}d we show the effect on the variance ratio when
changing the amplitude of the baryon-dark matter
streaming velocity. Solid (blue) lines correspond to $z=80$
and triple dot-dashed lines to $z=20$. In each group, top, middle and bottom
lines correspond to velocities $v_{bc}^{(bg)}=20,30$ and $40$km/s, respectively.
As expected, the larger the velocity, the stronger the effect of advection. 

More significant than the correction for advection is the 
extra power at small scales added by PBHs. This extra power
increases the rate of formation of matter halos at early times
(Kashlinsky 2016). To estimate the effect of PBHs on the formation
of halos in the presence of streaming motions, we will use
the Press-Schechter formalism (Press \& Schechter, 1974). 
The number density of objects $dn$
that are collapsing at redshift $z$ per unit of mass is a function
of the r.m.s. mass fluctuation $\sigma(M)$ and is given by
\begin{equation}\label{eq:ndensity}
\frac{dn}{dM}=\sqrt{\frac{2}{\pi}}\frac{\bar{\rho}_m}{M^2}
\frac{\delta_c}{\sigma_M}
\left|\frac{d\ln\sigma_M}{d\ln M}\right|\exp\left(-\frac{\delta_c^2}{2\sigma_M^2}\right)
\end{equation}
where $\bar{\rho}_m$ is the mean matter density, $\delta_c=1.686$ the critical 
overdensity above which a given mass fluctuation collapse and
\begin{equation}
\sigma_M^2(R)=\int_0^\infty \Delta_m^2(k)W^2(kR) d\ln k
\end{equation}
where $\Delta_m^2(k)$ is given in eq.~(\ref{eq:6}).
The variance of the matter density field $\sigma_M$
is computed on spheres of radius $R=(3M/4\pi\bar{\rho}_m)$ and
$W(x)=3j_1(x)/x$ is the Fourier transform of the top hat window function.

\begin{figure}
\centering 
\plotone{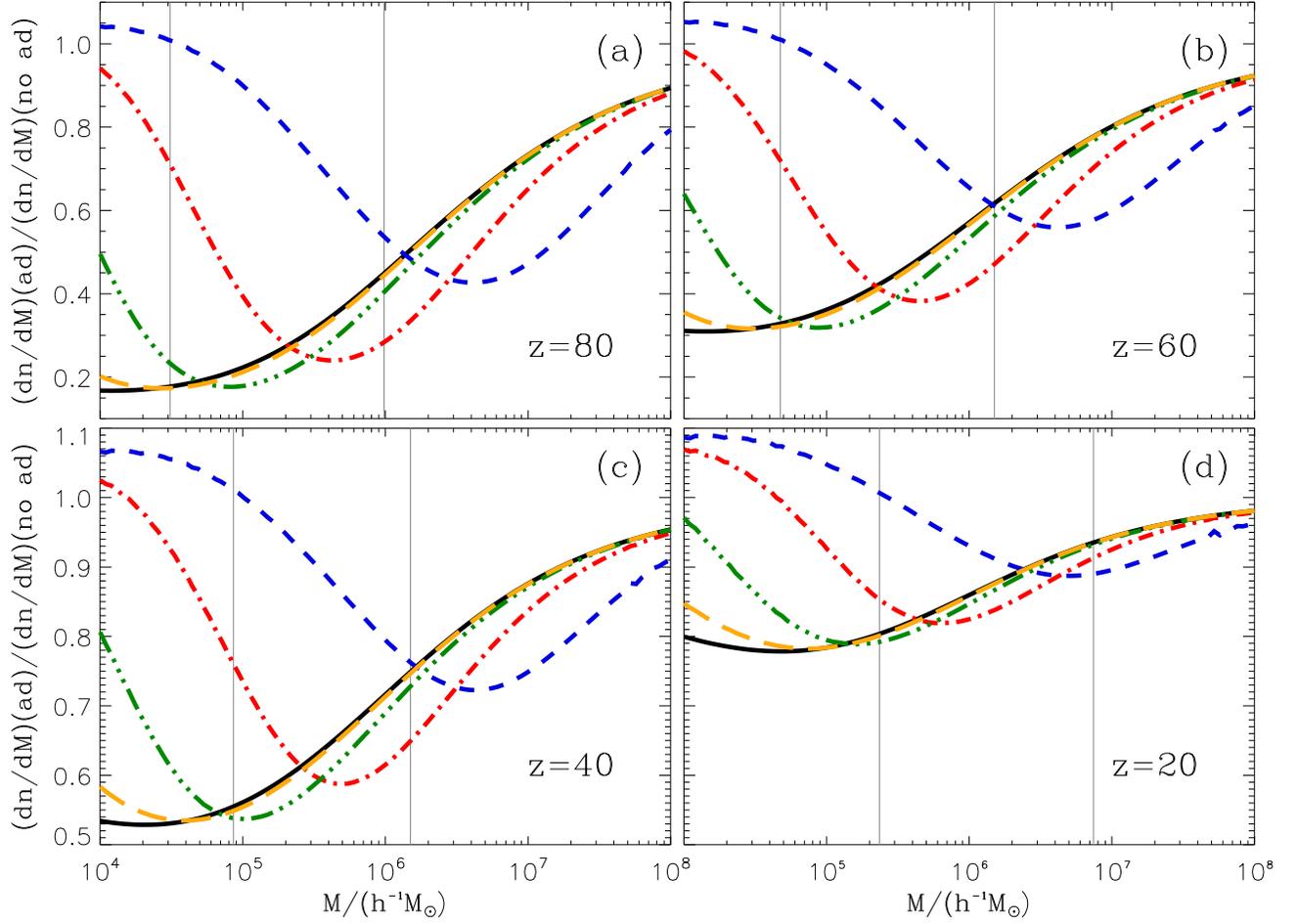}\bigskip
\caption{\small 
Ratio of the number density of collapsed objects per unit of  total mass for models
with advection to models without advection. (a) corresponds
to redshift $z=80$, (b) to $z=60$ and (c) to $z=40$ 
and (d) to $z=20$. Dashed (blue), dot-dashed (red), triple dot-dashed (green),
long dashed (gold) and solid (black) lines correspond to $f_{PBH}=
1,0.1,10^{-2},10^{-3},0$, respectively. The vertical gray lines correspond to
the masses of halos that have reached a virial temperature
$T_{vir}=10^3$K (left) and $T_{vir}=10^4$K (right).
}
\label{fig3}
\end{figure}

We compute the number density of objects in the range $[10^4,10^8]h^{-1}$M$_\odot$, 
where advection suppresses the formation of matter halos the most. 
Eq.~(\ref{eq:ndensity}) assumes that density perturbations are gaussian.
However PBHs are Poisson distributed and the density field
will not be gaussian, mainly at the scales where the PBH contribution
dominates. The power average in eq.~(\ref{eq:6}) will have
the same statistical distribution than $\Delta^2_m(k,\cos\theta)$. The
average will not dilute the non-gaussianity due to PBHs and, consequently
our treatment can only be considered an approximation.
Numerical simulations are needed to provide a more accurate estimate
of the effect of PBHs on the number density of matter halos.
Our results are presented in Fig.~\ref{fig3}. We plot the ratio of the
number densities (given by eq.~\ref{eq:ndensity}) of models with advection
to models without advection as a function of halo mass. The mass scale
is related to wavenumber as $k=2\pi(4\pi\bar{\rho}_m/3M)^{1/3}\approx
95(10^8h^{-1}M_\odot/M)^{1/3}h$Mpc$^{-1}$. The formation of halos at
$M\sim 10^4h^{-1}$M$_\odot$ are dominated by scales $k\sim 2h$kpc$^{-1}$.
As indicated, at those scales the non-linear evolution of PBHs damps the 
matter power spectrum (Inman \& Ali-Ha\"{\i}moud 2019). This non-linear effect
can only be estimated from simulations and we simplify the non-linear 
evolution by extrapolating the matter power spectrum linearly from 
its amplitude at $k=1h$kpc$^{-1}$ in the region $k\ge 1-10h$kpc$^{-1}$. 
Figs.~\ref{fig3}a-d correspond to redshifts
$z=80,60,40$ and $20$, respectively. Dashed (blue), dot-dashed
(red), triple dot-dashed (green), long-dashed (gold)
and solid (black) lines correspond to PBHs fractions
$f_{PBH}=(1,0.1,0.01,10^{-3},0)$. 
Comparing the results at these four redshifts,
we see that the effect of advection is diluted over time. This is expected
since baryons and DM tend to move together with decreasing redshift.

The small correction that PBHs induced in the streaming motion of
baryons is shown in Fig.~\ref{fig2}, combined with the larger values of $\sigma_8$
due to the extra power on small scales result on a large
increment on the number of low mass halos, as shown in  Fig.~\ref{fig3}.
At these small scales $\sigma(M)\sim 2-10$
and the exponential factor in eq.~(\ref{eq:ndensity}) is close to unity.
The abundance of objects is dominated by 
$\sigma_M^{-2}$ at a fixed mass. Since 
the variance of the density field with advection is smaller than without advection,
the number density of objects is larger, resulting in a significant
enhancement on the abundance of the less massive halos with respect to 
the standard model of DM particles. Fig.~\ref{fig3} shows
that in the considered redshift range, PBHs help the formation of halos 
with masses $M\le 10^{5-6}h^{-1}$M$_\odot$ in the presence of baryon advection. 
In the high mass end, the situation is reversed, with PBHs suppressing
the formation of halos. 
Fig.~\ref{fig3} also shows that when $f_{PBH}\sim 1$, objects in the range $M\sim 10^4$M$_\odot$
are formed more efficiently with than without advection.
This result can not be trusted due to our oversimplified treatment. 
First, as we have already indicated, numerical simulations showed
that non-linear evolution suppresses the isocurvature
component of the DM power spectrum 
(Inman \& Ali-Ha\"{\i}moud 2019), effect that we did not take into account.
Second, at these scales the PBHs dominate the matter power spectrum
and the Press-Schechter estimates are less accurate 
since density fluctuations depart from gaussianity.

Halos, made of PBHs, would be subject to stellar dynamical evolution 
effects leading to super-massive BHs early on (Kashlinsky \& Rees, 1983).
 Once formed, they would require efficient cooling mechanisms 
for baryons to collapse and form stars. Atomic cooling
is efficient when the halo reaches a virial temperature of $T_{vir}=10^4$,
while molecular hydrogen, when present, provides efficient cooling down
to temperatures $T_{vir}=300$K (Barkana \& Loeb 2001). In Fig.~\ref{fig3}
the vertical lines indicate the masses of matter halos with virial temperatures 
$T_{vir}=10^3$K, left and $10^4$K, right. In these estimates we have
assumed that the gas in those halos was primordial.
Since less massive halos are formed earlier and more abundantly, if
all physical conditions are equal, the collapse of low mass halos and the 
formation of first stars within them will be favored by the presence of PBHs.
Physically one expects that
if density perturbations grow faster in the absence of streaming motions,
it would also form objects more quickly. 
The interplay between the advection of baryons and the growth
of density perturbations at low mass scales facilitates the
formation of halos in the $10^4-10^5h^{-1}$M$_\odot$ mass range. Since 
the Press-Schechter formalism gives only an approximate picture of the 
collapse of non-linear
structures, numerical simulations are necessary to verify the correctness
of this prediction. Liu et al. (2022) included one simulation with PBHs and
streaming motions and found that the effect of advection
was weaker with PBHs, concluding that the perturbations from PBHs
accelerated the decoupling of baryons from the large scale flow.
Their model had $f_{PBH}=10^{-3}$ and the same 
BH masses and streaming velocities that in this work. Their simulation 
corresponds to the triple dot-dashed curves of Fig.~\ref{fig3} and their result
agrees with our prediction that PBHs speed up the
formation of objects in the range $(10^4-10^5)$M$_\odot$.

The suppression of small-scale baryonic structures due to
streaming velocities will affect the post-reionization gas
distribution, the gas temperature, the 21-cm line
temperature, the large scale distribution of stars and will 
alter the 21-cm signal. Its power spectrum will show
prominent Baryon Acoustic Oscillations (BAO) (Barkana 2016)
reflecting the BAO signature in Fig.~\ref{fig1}. The existence
of PBHs will further blur this picture since they will enhance
the formation of halos of mass $M\sim 10^4-10^5$M$_\odot$
while decreasing slightly more the formation of more
massive halos. The final effect will depend on the BH masses,
mass distribution and their fraction.

\section{Conclusions.}

We have analyzed the effect of PBHs in the evolution of small scale baryon and 
DM density perturbations in the presence of streaming motions. TH found that the
modes most strongly affected were those around $k\sim 200h$Mpc$^{-1}$.
PBHs add a constant isocurvature component to the matter density perturbations
that dominates the power at small scales. This extra power induces a
peculiar velocity field that does not disrupt the picture of the DM
moving at decoupling at supersonic speed with respect to baryons.
We have considered the power generated by PBHs of masses $M\sim 30$M$_\odot$
and with fractions $f_{PBH}=10^{-3}-1$ in the total DM budget.
We have found that PBHs favor the collapse of halos with mass $M\le 10^5-10^6$M$_\odot$ 
while slightly disfavor their formation for higher masses. 
We have quantified the amplitude of this effect, first predicted by 
Kashlinsky (2021). Since the less massive halos form earlier, the overall effect is
to weaken the role of streaming motions in the evolution of matter density
perturbation on Mpc scales, corroborating the result found by Liu et al. (2022)
in their numerical simulations. 

We have predicted that PBHs 
help halos of masses $10^4-10^5$M$_\odot$ to collapse slightly more effectively
in the presence of advection than in their absence. This shows the limitations
of our model since in addition to non-linear effects and the non-gaussianity of the
PBH power spectrum, we are not taking into account
variations on the physical conditions of the gas, the cooling process
or the feedback from
PBH accretion. All those effects can only be verified with numerical simulations
since the overall picture is unlikely to be as simple as described
by our analytical model. If the ratio 
of PBHs to the total density of DM is small, one would expect PBHs
to accrete large amounts of matter but contributing little to the 
matter power spectrum. When all the DM is made of BHs, the 
Poisson fluctuation in the PBHs number density would dominate
the dynamics on small scales. Both regimes, termed "seed" and "Poisson" limits 
(Carr \& Silk 2018) can be explored by analytical
methods. For PBHs of masses $10-100$M$_\odot$ and fractions
$f_{PBH}=10^{-3}-0.1$ DM structures around individual PBHs will interact and
their dynamics can only be followed with time consuming
N-body simulations as in Inman \& Ali-Ha\"{\i}moud (2019).
PBHs shift star formation to more massive 
halos, accelerating structure formation more strongly 
in regions with higher initial overdensities (Liu et al. 2022),
physics that is not captured by our formalism. Still, analytical
methods like the one presented here provide a general understanding of
the physical processes without expensive computations.

\bigskip
{\bf Acknowledgments:}
We thank A. Kashlinsky for discussions and we acknowledge financial support from 
grants PGC2018-096038-B-I00 (MINECO and FEDER, 
"A way of making Europe") and SA083P17 from the Junta de Castilla y Le\'on.

\clearpage \pagestyle{plain}

\end{document}